# Deuterium enrichment of ammonia produced by surface N+H/D addition reactions at low temperature

G. Fedoseev[1, a)], S. Ioppolo[2, 3] and H. Linnartz[1]

[1]*Sackler Laboratory for Astrophysics, Leiden Observatory, University of Leiden, PO Box 9513, NL 2300 RA Leiden, The Netherlands*

[2]*Division of Geological and Planetary Sciences, California Institute of Technology, 1200 E. California Blvd., Pasadena, California 91125, USA*

[3]*Institute for Molecules and Materials, Radboud University Nijmegen, PO Box 9010, NL 6500 GL Nijmegen, The Netherlands*

**ABSTRACT**

The surface formation of $NH_3$ and its deuterated isotopologues - $NH_2D$, $NHD_2$, and $ND_3$ - is investigated at low temperatures through the simultaneous addition of hydrogen and deuterium atoms to nitrogen atoms in CO-rich interstellar ice analogues. The formation of all four ammonia isotopologues is only observed up to 15 K, and drops below the detection limit for higher temperatures. Differences between hydrogenation and deuteration yields result in a clear deviation from a statistical distribution in favour of deuterium enriched species. The data analysis suggests that this is due to a higher sticking probability of D atoms to the cold surface, a property that may generally apply to molecules that are formed in low temperature surface reactions. The results found here are used to interpret ammonia deuterium fractionation as observed in pre-protostellar cores.

**1 INTRODUCTION**

The characterization of different evolutionary stages of star formation is essential to understand the origin of molecular complexity in space. The ratio of a deuterated species over its counterpart containing H, *i.e.*, the deuterium fractionation, is known to be a good tool to discriminate between different processes taking place along this evolutionary trail. The D/H ratio in the interstellar medium (ISM) is $\sim 1.5 \cdot 10^{-5}$. In cold dense cores ($T \sim 10\text{-}20$ K and $n \sim 10^6$ cm$^{-3}$), the depletion of C-bearing molecules leads to an enhancement of the deuterium fractionation, because the $H_2D^+$ ion, the gas-phase progenitor of most of the deuterated species, is not destroyed by CO (*e.g.*, Roberts & Millar 2006, Bacmann et al. 2003, Pillai et al. 2007). Apart from gas-phase reactions, also grain surface chemistry enhances molecular D/H ratios under cold dense cloud conditions: *e.g.*, via low-temperature surface reactions, ion- and photo-dissociation of solid species, and thermally induced exchange reactions on icy grains (*e.g.*, Tielens 1983, Brown & Millar 1989a, 1989b). The idea that such processes lead to deuterium enrichment is usually based on the 'zero-energy argument', *i.e.*, the assumption that species with D atoms have a lower zero-energy and, therefore, are more stable and more likely to be formed over their H-atom counterparts at low temperatures (10-20 K). Moreover, if processes are recurrent in space, then deuterium enrichment can be further enhanced over a number of

---

a) Author to whom correspondence should be addressed. Electronic mail: fedoseev@strw.leidenuniv.nl

cycles (Tielens 1983). During the formation stage of a protostar, the D/H value is found to decrease again because of the strong UV field and shocks in the outflows caused by the newborn star (Crapsi et al. 2005, Emprechtinger et al. 2009). Also much later, during the planetary stage, D/H ratios are important. Recent observations with the *Herschel* Space Observatory have revealed that although the D/H mean value in Oort-cloud comets is $\leq 3 \cdot 10^{-4}$, the D/H ratio in Jupiter-family comets is very close to the Vienna standard mean ocean water (VSMOW) value of $\sim 1.5 \cdot 10^{-4}$. This result seems to suggest that a significant delivery of cometary water to the Earth-Moon system occurred shortly after the Moon-forming impact (Hartogh et al. 2011, Bockelée-Morvan et al. 2012, Lis et al. 2013). In this scenario, also other complex prebiotic species may have been delivered to Earth following the same route of water molecules.

To date, a conspicuous number of molecules detected in the ISM, including methanol, water and ammonia exhibit deuterium enrichment. Laboratory experiments show that all these species are formed on the surface of icy dust grains mostly through hydrogenation reactions at low temperatures. In particular, the presence of species like $H_2CO$ and $CH_3OH$ (CO+H), $NH_3$ (N+H), $CH_4$ (C+H), $H_2O$ ($O/O_2/O_3$+H), and possibly also $NH_2OH$ (NO+H, $NO_2$+H) is (largely) explained through sequential H-atom addition to various precursors (*e.g.*, Hiraoka et al. 1994, Hiraoka et al. 1995, Hiraoka et al. 1998, Zhitnikov & Dmitriev 2002, Watanabe & Kouchi 2002, Miyauchi et al. 2008, Ioppolo et al. 2008, Mokrane et al. 2009, Congiu et al 2012, Ioppolo et al 2014). As a consequence, surface deuteration reactions will be at play as well, contributing with different efficiencies. Indeed, laboratory work shows that different molecules undergo different surface deuterations (Nagaoka at al 2005, Ratajczak et al 2009, Hidaka at al 2009, Weber et al 2009, Kristensen et al 2011). For example, the formation of $H_2O$/HDO through $OH+H_2/D_2$ and $H_2O_2$+H/D (Oba et al 2012, Oba et al 2014) shows a preference for hydrogenation that has been explained by a higher quantum tunneling efficiency. In fact, the different transmission mass involving abstraction/addition of hydrogen atoms over similar reactions with deuterium is assumed to cause higher hydrogenation rates. On the other hand, Nagaoka at al. (2005), Nagaoka (2007), and Hidaka at al. (2009) demonstrated that hydrogen atoms can be abstracted from methanol and its isotopologues and substituted by D atoms upon D-atom exposure of solid $CH_3OH$, $CH_2DOH$, and $CHD_2OH$. However, H-atom exposure of $CD_3OH$, $CD_2HOH$, and $CDH_2OH$ does not result in abstraction or substitution of D atoms with H atoms. Therefore, (partially) deuterated methanol is expected to be enriched in space. These examples make clear that a straight forward interpretation of H/D ratios in the ISM is far from trivial. Nevertheless, understanding the mass dependence of all possible processes helps in pinning down the relevant reactions taking place.

The present laboratory work focuses on the competition between hydrogenation and deuteration during the solid state formation of ammonia. It is commonly believed that $NH_3$ can be formed both in the gas phase - through a series of ion-molecule reactions (see Herbst & Klemperer 1973, Scott et al. 1997) - and in the solid state - through three sequential H-atom additions to a single nitrogen atom on the surface of an interstellar ice grain. Gas-phase observations toward 'pre-protostellar cores' show an increased D/H fractionation of $NH_3$ (Hatchell 2003, Busquet et al. 2010). $NH_2D$ and $NHD_2$ are also detected in the dark cloud L134N and in IRAS 16293E (Rouef et al. 2000, Loinard et al. 2001). In these sources, the $NH_3$:$NH_2D$:$NHD_2$ gas-phase abundances (1:0.1:0.005 and 1:0.07:0.03, respectively) are orders of magnitudes higher than the ratios expected from cosmic D/H abundances. To date, there is no direct astronomical observation of the D/H ratio for ammonia ice, but the astronomical gas-phase data may actually reflect the solid state deuterium enrichment as well. For instance, observations toward the shocked region L1157-B1 provide evidence for a chemical enrichment of the interstellar gas by the release of dust ice mantles and show an indirect upper limit for $NH_2D/NH_3$ of $3 \cdot 10^{-2}$ (Codella et al. 2012).

Recently, several groups studied the surface hydrogenation of N atoms (see Hiraoka et al. 1995, Hidaka et al. 2011, Fedoseev et al. 2014). It is clear that a better understanding of the D/H fractionation mechanism of this process is required to interpret the gas-phase observations of deuterium enriched $NH_3$. Moreover, both Hidaka et al. (2011) and Fedoseev et al. (2014) – see the accompanying paper - conclude experimentally that the formation of $NH_3$ by hydrogenation of N atoms involves a Langmuir-Hinshelwood (L-H) mechanism. Therefore, key parameters for the formation rates of $NH_3:NH_2D:NHD_2:ND_3$ during hydrogenation/deuteration of N atoms, are the sticking coefficients for H and D atoms, the activation energies needed for surface diffusion and desorption and the reaction barriers for interactions with other species. Thus, the deuterium enrichment of ammonia offers a diagnostic tool to investigate the L-H mechanism. In this paper the astronomical implications and details on the physical-chemical properties of the deuterium enrichment of ammonia are discussed.

## 2 EXPERIMENTAL PROCEDURE

### 2.1 Experimental setup

The experiments are performed under ultra-high vacuum (UHV) conditions, using our SURFRESIDE$^{(2)}$ setup, first in a one-atom beam, and since 2012 in a double-atom beam configuration. The latter setup allows for the simultaneous use of two atom-beams together with regular molecular dosing lines as described in the accompanying paper. Further details of the original and extended setups are available from Fuchs et al. (2009) and Ioppolo et al. (2013), respectively.

SURFRESIDE$^2$ consists of three distinct UHV sections, including a main chamber and two atom-beam line chambers. Shutters separate the beam lines from the main chamber and allow for an independent operation of the individual parts. Two different atom sources are implemented: a Hydrogen Atom Beam Source (HABS, Dr. Eberl MBE-Komponenten GmbH, see Tschersich 2000) that produces H or D atoms through thermal cracking of $H_2$ or $D_2$; and a Microwave Atom Source (MWAS, Oxford Scientific Ltd, see Anton et al. 2000) that generates H, D, O, and, specifically for this work, N atoms using a microwave discharge (300 W at 2.45 GHz). $H_2$ (Praxair 5.0), $D_2$ (Praxair 2.8), and $N_2$ (Praxair 5.0) are used as precursor gasses. A nose-shaped quartz pipe is placed after each shutter along the path of the atom beam to efficiently quench the excited electronic and ro-vibrational states of the formed atoms and non-dissociated molecules through collisions with the walls of the pipe before they reach the ice sample. The geometry is designed in such a way that this is realized through at least four wall collisions before atoms can leave the pipe. In this way, 'hot' species cannot reach the ice. All atom fluxes are in the range between $10^{11}$ and $10^{13}$ atoms cm$^{-2}$ s$^{-1}$ at the substrate position. The calibration procedures are described in Ioppolo et al. (2013).

In the main chamber, ices are deposited with monolayer precision (where 1 ML = $10^{15}$ molecules cm$^{-2}$) at astronomically relevant temperatures (starting from 13 K and upwards) onto a 2.5 x 2.5 cm$^2$ gold substrate. The substrate is mounted on the tip of a cold head and full temperature control is realized using a Lakeshore temperature controller. The absolute temperature precision is ~2 K, and the relative precision between two experiments is below 0.5 K. Two additional dosing lines are implemented to allow for a separate deposition of stable molecules. In this way, it becomes possible to co-deposit atoms and molecules and to simulate various molecular environments as typical for different evolutionary stages in interstellar ices: *e.g.*, elementary processes in polar ($H_2O$-rich) or non-polar (CO-rich) interstellar ice analogues (see Ioppolo et al. 2014). Pre-deposited ices can also be studied with this system. Reflection

Absorption Infrared Spectroscopy (RAIRS) and Temperature Programmed Desorption (TPD) in combination with a quadrupole mass spectrometer (QMS) are used as analytical tools to characterize the ice composition both spectroscopically and mass spectrometrically. Since $NH_3$ and $ND_3$ molecules can participate in thermally induced deuterium exchange reactions, RAIRS is used as main diagnostic tool, complemented with TPD data.

## 2.2 Performed experiments

Two sets of conceptually different (control) experiments are performed. The first set focuses on H/D exchange reactions with pre- and co-deposited $NH_3$ with D atoms. The second set, which is the core of this work, deals with the isotopic fractionation in sequential H- and D-atom reactions with atomic nitrogen in co-deposition experiments:

1. $NH_3$ + D studies are performed using both pre- and co-deposition experiments. $NH_3$ (Praxair 3.6) and $D_2$ are prepared in distinct pre-pumped (< $10^{-5}$ mbar) all-metal dosing lines. The pre-deposition of $NH_3$ ice is performed under an angle of 45° and with a controllable rate of 4.5 ML min$^{-1}$ on a 15 K gold substrate. During deposition, RAIR difference spectra are acquired every minute with respect to a pre-recorded spectrum of the bare gold substrate. After $NH_3$ deposition, a new background reference spectrum is acquired, and the pre-deposited $NH_3$ ice is exposed to a constant flux of D atoms normal to the surface of the sample. RAIR difference spectra are acquired every 10 minutes to monitor the ice composition *in situ*. During the co-deposition experiment, $NH_3$ and D atoms are deposited simultaneously on the gold substrate with a constant rate. Also here, RAIR difference spectra are acquired every 10 minutes with respect to a spectrum of the bare substrate.

2. Significantly more complex from an experimental point of view is the study of H/D fractionation in ammonia isotopologues formed upon co-deposition of hydrogen, deuterium, and nitrogen atoms. This experiment can be performed by co-depositing H and D atoms generated in the HABS with N atoms from the MWAS. A mixed H- and D-atom beam is obtained by thermal cracking a $H_2:D_2$=1:1 gas mixture. The mixture is prepared by filling up two independent pre-pumped (< $10^{-5}$ mbar) parts of the dosing line with a known volume-to-volume ratio of $H_2$ and $D_2$ gasses. The gasses are subsequently allowed to mix in the total volume of the dosing line. The pre-mixed gas is then introduced in the thermal cracking source using a precise leak valve to control the gas flow. In a similar way, a $N_2$ gas line is used as input for the microwave plasma source. Another pre-pumped (< $10^{-5}$ mbar) dosing line is used to deposit CO under the same experimental conditions as previously described in the accompanying paper by Fedoseev et al. (2014). As explained there the addition of CO simulates a more realistic interstellar ice environment. To guarantee stable operational conditions, both thermal cracking and microwave plasma sources are operated ('backed') for at least 30 minutes prior to co-deposition. A simultaneous co-deposition experiment is then performed at the desired sample temperatures (13-17 K) by using all three H-, D-, and N-atom beams as well as a molecular CO beam for a period of typically four hours. During this co-deposition experiment, RAIR difference spectra are acquired every 5 minutes with respect to a spectrum of the bare gold substrate. All relevant experiments are summarized in Table 1.

Here, the main challenge is to evaluate with the highest possible precision the H:D ratio in the mixed atom beam to allow for a quantitative study of the competition between hydrogenation and deuteration of ammonia ice. Ioppolo et al. (2013) characterized the atom fluxes produced by SURFRESIDE$^2$ and found that, for identical settings and with $H_2$ and $D_2$ used separately, the resulting H-atom fluxes at the surface are a factor of ~2 higher than the measured D-atom fluxes. However, this does not necessarily mean that using a $H_2:D_2 = 1:1$ mixed gas will result in a production of H and D atoms in a 2:1 ratio. This is because differences in the $H_2$ and $D_2$ flow rates through the leaking valve as well as different recombination efficiencies of H+H, D+D, and H+D on the walls of the quenching quartz pipe can change the final H:D ratio. Moreover, the absolute flux calibrations come with large uncertainties (~30-50%).

Therefore, after each H:D:N experiment, we performed a control co-deposition experiment of $H:D:O_2$ ~ 1:1:100 at the same temperature of the H:D:N experiment to derive the effective H and D fluxes at the substrate surface for all the different temperatures investigated. These control experiments lead to the formation of two products: $HO_2$ and $DO_2$ through the reactions $H/D + O_2 = HO_2/DO_2$. The surface formation of $HO_2$ and $DO_2$ has been the topic of several studies and is known to be a nearly barrierless process (*e.g.*, Cuppen et al. 2010). Therefore, the total amount of final products and the $HO_2:DO_2$ ratio are expected to be independent of temperature - as we verified here in the range between 13 and 17 K - and reflect the effective H- and D-atom fluxes at the surface. Band strengths of selected mid-IR vibrational modes are needed to quantify the final abundances of $HO_2$ and $DO_2$ formed in our control experiments. Unfortunately, there is no experimental data on $HO_2$ and $DO_2$ band strengths in an $O_2$ matrix. Therefore, we used integrated band areas to calculate the ratio between OD:OH stretching modes and the ratio between DOO:HOO bending modes for all the investigated temperatures. The two ratios give the following values of 1.48 and 1.47, respectively, which are temperature independent.

To quantitatively link the OD:OH and DOO:HOO ratios as obtained from the integrated band areas with the ratio of the final amount of $HO_2$ and $DO_2$, we still need to know at least the band strength ratios for the selected vibrational modes. Therefore, we assume that the band strength ratio between the DOO and the HOO bending modes of $DO_2$ and $HO_2$, respectively, is the same as for the band strength ratio between the DOO and the HOO bending modes of $H_2O_2$ and $D_2O_2$ obtained in $O_2$ + H/D experiments by Miyauchi et al. (2008) and Oba et al. (2014). As shown in Oba et al. (2014), the peak positions of $H_2O_2$ (1385 cm$^{-1}$) and $D_2O_2$ (1039 cm$^{-1}$) are similar to the peak positions of $HO_2$ (1391 cm$^{-1}$) and $DO_2$ (1024 cm$^{-1}$) in our $O_2$ matrix experiments (Bandow & Akimoto 1985). The DOO:HOO band strength ratio in Miyauchi et al. (2008) and Oba et al. (2014) is determined as 1.4. We also assume that the OD:OH band strength ratio from the stretching modes of $HO_2$ and $DO_2$ is the same as for the OD:OH band strength ratio from the stretching modes of $H_2O$ and $D_2O$ determined by Bergren et al. (1978) and Miyauchi et al. (2008). In this case, an OD/OH band strength ratio of 1.5 was found. By scaling the integrated band area ratios of the two selected vibrational modes (OD:OH = 1.48 and DOO:HOO = 1.47) for their respective band strength ratios (OD:OH = 1.5 and DOO:HOO = 1.4) we conclude that the ratio of our effective H- and D-atom fluxes is equal to one within the experimental uncertainties.

Once the first $H:D:O_2$ ~ 1:1:100 control experiment is completed, the ice is usually sublimated and a second control co-deposition experiment of H and D atoms with $O_2$ molecules is repeated always at 13 K. This second control experiment is meant to monitor the day-to-day reproducibility of the H- and D-atom fluxes. We did not find any major fluctuations in the H:D-atom beam over the course of our experiments.

**Table 1.** List of the performed experiments.

| Ref. N | Experiment | Method[a] | Ratio | $T_{sample}$ (K) | $R_{dep}$ (ML min$^{-1}$) | Atom-flux$^{TL}$ ($10^{15}$ cm$^{-2}$ min$^{-1}$)[b] | Atom-flux$^{PL}$ ($10^{15}$ cm$^{-2}$ min$^{-1}$)[b] | $t_{total}$ (min) | TPD |
|---|---|---|---|---|---|---|---|---|---|
| SURFRESIDE | | | | | | | | | |
| **Isotopic exchange in NH$_3$+D system** | | | | | | | | | |
| | | | | | NH$_3$ | D (from D$_2$) | | | |
| 1.1 | NH$_3$:D | pre-dep | - | 15 | 4.5 (50 ML) | 0.7 | - | 120 | - |
| | | | | | | 2.5 | - | 60 | QMS |
| 1.2 | NH$_3$:D | co-dep | 1:5 | 15 | 0.5 | 2.5 | - | 120 | - |
| SURFRESIDE$^2$ | | | | | | | | | |
| **Deuterium fractionation of ammonia isotopologues produced by simultaneous surface (H+D)-atom addition to N-atoms** | | | | | | | | | |
| | | | | | CO | H+D (H$_2$:D$_2$ = 1:1) | N (from N$_2$) | | |
| 2.1 | N:(H+D):N$_2$:CO | co-dep | 1:15$^c$:100:100 | 13 | 0.5 | 0.075$^c$ | 0.005 | 180 | QMS$^{1K/5K}$ |
| 2.2 | N:(H+D):N$_2$:CO | co-dep | 1:15$^c$:100:100 | 13 | 0.5 | 0.075$^c$ | 0.005 | 240 | - |
| 2.3 | N:(H+D):N$_2$:CO | co-dep | 1:15$^c$:100:100 | 14 | 0.5 | 0.075$^c$ | 0.005 | 240 | - |
| 2.4 | N:(H+D):N$_2$:CO | co-dep | 1:15$^c$:100:100 | 15 | 0.5 | 0.075$^c$ | 0.005 | 240 | - |
| 2.5 | N:(H+D):N$_2$:CO | co-dep | 1:15$^c$:100:100 | 16 | 0.5 | 0.075$^c$ | 0.005 | 240 | - |
| 2.6 | N:(H+D):N$_2$:CO | co-dep | 1:15$^c$:100:100 | 17 | 0.5 | 0.075$^c$ | 0.005 | 240 | - |
| | | | | | CO | H (from H$_2$) | N (from N$_2$) | | |
| 3.1 | N:H:N$_2$:CO | co-dep | 1:20:100:100 | 13 | 0.5 | 0.1 | 0.005 | 90 | - |
| | | | | | CO | D (from D$_2$) | N (from N$_2$) | | |
| 3.2 | N:D:N$_2$:CO | co-dep | 1:20:100:100 | 13 | 0.5 | 0.1 | 0.005 | 90 | - |
| **Determination of H:D atom ratio in produced mixed (H+D)-atom beam** | | | | | | | | | |
| | | | | | O$_2$ | H+D (H$_2$:D$_2$ = 1:1) | | | |
| 4.1 | (H+D):O$_2$ | co-dep | 1:600 | 13 | 45 | 0.075$^c$ | - | 10 | - |
| 4.2 | (H+D):O$_2$ | co-dep | 1:600 | 14 | 45 | 0.075$^c$ | - | 10 | - |
| 4.3 | (H+D):O$_2$ | co-dep | 1:600 | 15 | 45 | 0.075$^c$ | - | 10 | - |
| 4.4 | (H+D):O$_2$ | co-dep | 1:600 | 16 | 45 | 0.075$^c$ | - | 10 | - |
| 4.5 | (H+D):O$_2$ | co-dep | 1:600 | 17 | 45 | 0.075$^c$ | - | 10 | - |

[a]Experiments are performed by pre-depositon (pre-dep) and co-deposition (co-dep) techniques; $R_{dep}$ is the deposition rate of a selected molecule expressed in ML min$^{-1}$ under the assumption that 1 L (Langmuir) exposure leads to the surface coverage of 1 ML; $T_{sample}$ is the substrate temperature during co-deposition; Atom-flux$^{TL}$ is the HABS atom flux; Atom-flux$^{PL}$ is the plasma cracking line atom flux; $t_{total}$ is the total time of co-deposition; TPD is the temperature programmed desorption experiment performed afterward with the TPD rate indicated.
[b]Absolute uncertainties of H/D- and N- fluxes are 50 and 40%, respectively; the relative uncertainty between two of the same N+H/D experiments is <20%, and is as low as a few percent for H/D+O$_2$ experiments;
[c]Since the exact atom-beam composition for H$_2$:D$_2$ = 1:1 feeding mixture is unknown, the absolute atom flux corresponding to the sum of the two distinct fluxes (one for pure H$_2$ and one for pure D$_2$) divided by 2 is used to present co-deposition ratios and total fluences.

## 3 RESULTS AND DISCUSSION

### 3.1 Deuterium exchange in NH$_3$ + D system

Nagaoka et al. (2005) and Hidaka et al. (2009) found that both solid H$_2$CO and CH$_3$OH participate in abstraction reactions with D atoms in the 10-20 K temperature range yielding D-substituted methanol isotopologues in H$_2$CO+D and CH$_3$OH+D experiments. In the same work, Nagaoka et al. (2005) investigated similar reactions for the NH$_3$+D system and reported that no abstraction is observed in the exposure of pre-deposited NH$_3$ to cold D atoms at temperatures below 15 K. In the present work, we verify this conclusion for our experimental pre-deposition conditions and we further expand on this by performing co-deposition experiments.

In the top panel of Figure 1, two RAIR spectra are presented: (a) a spectrum obtained after 50 ML deposition of pure NH$_3$ ice, and (b) the RAIR difference spectrum obtained after exposure of this ice to $8\cdot10^{16}$ D atoms cm$^{-2}$. One can see that all four main spectroscopic absorption features of NH$_3$ ($\nu_1$ = 3217 cm$^{-1}$, $\nu_2$ = 1101 cm$^{-1}$, $\nu_3$ = 3385 cm$^{-1}$, $\nu_4$ = 1628 cm$^{-1}$, see Reding & Hornig 1954) are visible in the spectra of pure NH$_3$ ice. It should be noted that the peak positions are significantly shifted in this pure ammonia environment from the ones in rare gas matrix-isolated NH$_3$ due to the presence of hydrogen bonds between NH$_3$ molecules (H$_2$N-H···NH$_3$) (compare for example data in Reding & Hornig 1954 with those in Abouaf-Marguin et al. 1977 and Hagen & Tielens 1982). Spectrum (b), in turn, shows that none of the NH$_2$D absorption bands demonstrates a noticeable growth upon D-atom exposure, at least not within our experimental detection limits. A small feature in the N-H/O-H stretching region (3000-3500 cm$^{-1}$), and a shift of the peak centred at 1100 cm$^{-1}$ (see the negative bump in spectrum b) is likely due to background water deposition on top of $NH_3$ during the 2 hours of D-atom exposure. It is known that position and strength of $NH_3$ absorption bands can vary in presence of $H_2O$ ice, since new hydrogen bonds are formed with $H_2O$ molecules, *i.e.*, HOH···$NH_3$, $H_2O$···$HNH_2$ (Bertie & Shehata 1985).

In order to further increase the sensitivity of our experimental technique, co-deposition experiments of $NH_3$ molecules with D atoms are performed. With a sufficiently high number of D atoms over $NH_3$ molecules (*e.g.*, 5:1 ratio), virtually, every deposited $NH_3$ molecule is available for reaction with D atoms, and a greater amount of products may be formed and be available for detection by RAIRS. In the lower panel of Figure 1, the results of this experiment are shown; there is no observable difference between the $NH_3$+D and pure $NH_3$ deposition experiments at 15 K, and none of the possible D-substituted $NH_3$ isotopologues, *i.e.,* $NH_2D$, $NHD_2$, $ND_3$, can be spectroscopically identified. This further constrains the conclusion made by Nagaoka et al. (2005) that the reaction $NH_3$+D does not take place at temperatures lower than 15 K. Moreover, because of the aforementioned 'zero-energy argument', we expect that also the reaction $ND_3$+H is not efficient at low temperatures.

It should be noted that hydrogen bonds between $NH_3$ molecules stabilise $NH_3$ and make abstraction reactions by D atoms thermodynamically less favourable (in a way similar to the $H_2O$+D case). Therefore, performing the $NH_3$ + D experiments in a non-polar matrix could potentially prevent the hydrogen bond formation and in return help to overcome the activation barrier for the exchange reaction at low temperatures. Although this is an interesting project, we decided to perform another challenging set of experiments with the intent to study the deuterium fractionation during ammonia formation. The outcome of these experiments is presented in the next section.

8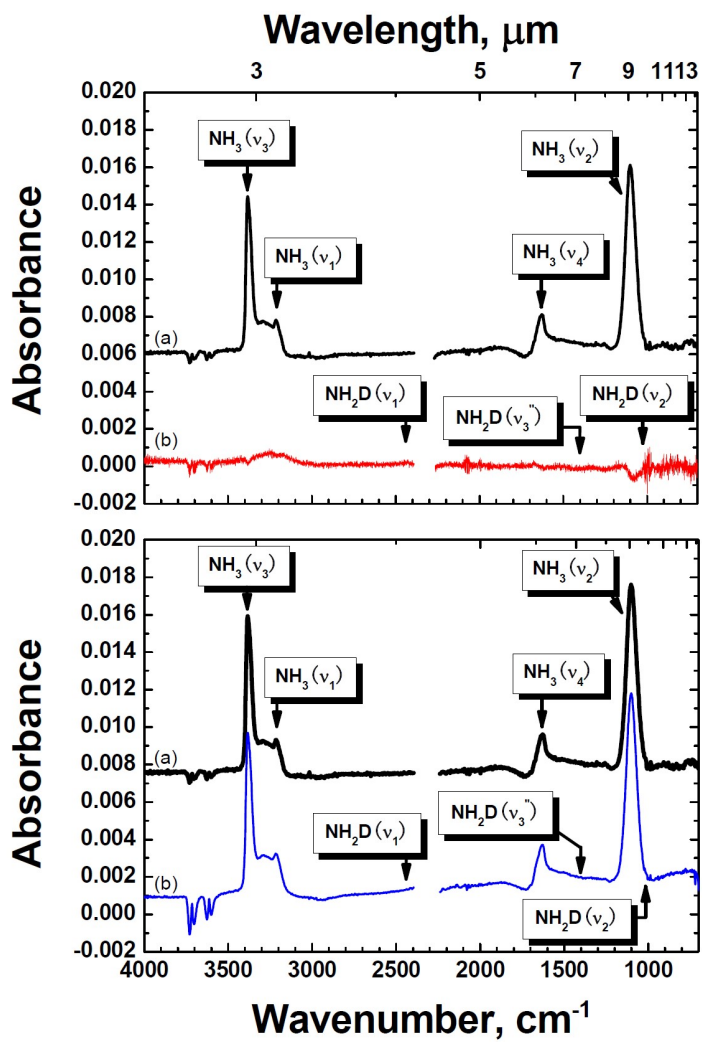

**Figure 1.** Upper panel: (a) RAIR spectrum of 50 ML of pure NH$_3$ at 15 K (experiment 1.1); (b) RAIR difference spectrum obtained after exposure of 50 ML of pre-deposited NH$_3$ ice with 8·10$^{16}$ D atoms cm$^{-2}$ at 15 K (experiment 1.1). Lower panel: (c) simultaneous co-deposition of 50 ML of NH$_3$ with 3·10$^{17}$ D atoms cm$^{-2}$ at 15 K, where D:NH$_3$ = 5:1 (experiment 1.2). The removed window between 2250 and 2380 cm$^{-1}$ and the region from 3550 to 3650 cm$^{-1}$ present absorption bands due to atmospheric CO$_2$ along the path of the IR beam outside the UHV setup.

## 3.2 H/D fractionation of ammonia (isotopologues) produced by hydrogenation of N atoms at low temperatures

Here, we show the first experimental results for deuterium fractionation in NH$_3$ molecules during the formation of ammonia through N-atom hydrogenation/deuteration in interstellar ice analogues. A number of experiments has been performed in which N atoms are co-deposited at 13 K with CO molecules and a mixture of H and D atoms in a single beam (see Table 1 experiments 2.1-3.2). Now, all four possible ammonia D-isotopes can be identified following the identification of several spectral features: NH$_3$ ($\nu_2$ = 975 cm$^{-1}$ and $\nu_4$ = 1625 cm$^{-1}$); NH$_2$D ($\nu_2$ = 909 cm$^{-1}$ and $\nu_4$ = 1389 cm$^{-1}$); NHD$_2$ ($\nu_2$ = 833 cm$^{-1}$, $\nu_4$ =

1457 cm$^{-1}$, and $\nu_3$ = 2551 cm$^{-1}$); and ND$_3$ ($\nu_2$ = 762 cm$^{-1}$ and $\nu_3$ = 1187 cm$^{-1}$). Assignments are taken from Reding & Hornig (1954), Abouaf-Marguin et al. (1977), Nelander (1984), and Koops et al. (1983). Figure 2a shows the RAIR spectrum of a co-deposition of N:(H+D):N$_2$:CO = 1:15:100:100 at 13 K with a H$_2$/D$_2$ = 1 ratio and the N$_2$ is due to the undissociated precursor gas. In this experiment, also HDCO can be observed at $\nu_2$ = 1708 cm$^{-1}$ and $\nu_3$ = 1395 cm$^{-1}$ (Hidaka et al. 2009), while both H$_2$CO and D$_2$CO are present only as trace signals at 1733 cm$^{-1}$ and 1682 cm$^{-1}$, respectively.

In order to make the interpretation of our data quantitative, absolute band strengths are needed. For NH$_3$ and ND$_3$ absolute intensity measurements exist. There are no available data for the band strengths of the partially deuterated ammonia isotopes - NH$_2$D and NHD$_2$ - for which only predictions have been reported so far (Koops et al. 1983). Since in our study $\nu_2$ is the only mode that is present for all ammonia isotopologues and this vibrational mode can be affected by the ice lattice, we have decided to apply a different method for the quantitative characterization of ammonia deuterium fractionation at low temperatures. In the accompanying paper (Fedoseev et al. 2014), we have already shown that under our experimental conditions the amount of produced ammonia is determined by the amount of nitrogen atoms available for hydrogenation. In the case of a pure statistical (*i.e.*, mass independent) distribution of the products of simultaneous H- and D-atom additions to nitrogen atoms, the final yield distribution of NH$_3$:NH$_2$D:NHD$_2$:ND$_3$ should be about 1:3:3:1, assuming a H:D = 1:1 ratio, as determined in the experimental section.

Therefore, in Figure 2 we compare the spectrum from a co-deposition of N:(H+D):N$_2$:CO = 1:15:100:100 at 13 K (Fig. 2a) with spectra from a nitrogen-atom co-deposition experiment with only H atoms (Fig 2b) or only D atoms (Fig 2c), in such a way that the total amount of deposited N atoms in the experiment shown in Fig. 2a is 8 times higher than that shown in each of the two other spectra. It is illustrated in Fedoseev et al. (2014) that under our experimental conditions a full conversion of N atoms into the final product (NH$_3$) is achieved. Therefore, if our assumption of a 1:3:3:1 distribution of the formed isotopologues is correct, Fig. 2b should represent a statistical weight of the formed NH$_3$, and Fig. 2c of ND$_3$. In this distribution, the amount of produced NH$_3$ is 8 times lower than the total amount of all the formed ammonia isotopologues, *i.e.*, only 1/8 part of the deposited N atoms should be converted to NH$_3$. The same applies to ND$_3$. The comparison of the NH$_3$ and ND$_3$ band areas in Fig. 2a with the NH$_3$ band area in Fig. 2b and the ND$_3$ band area in Fig. 2c shows that, for the chosen settings, there is a deviation from the statistical 1:3:3:1 distribution in favor of an increase in the production of deuterated species. In particular, by comparing the total amount of ND$_3$ in Figs. 2a and 2c, we estimate that every deuteration reaction has a probability of almost a factor 1.7 higher to occur over the corresponding hydrogenation reactions. The area of ND$_3$ in Fig. 2a is two times larger than the ND$_3$ area in Fig. 2c. To achieve such enhancement of ND$_3$ production over three subsequent additions of H or D atoms to an N atom, every deuteration reaction should have a probability of a factor 1.7 higher to occur, resulting in a final NH$_3$:NH$_2$D:NHD$_2$:ND$_3$ distribution of 0.4:2.1:3.5:2. This is further discussed in the following sections.

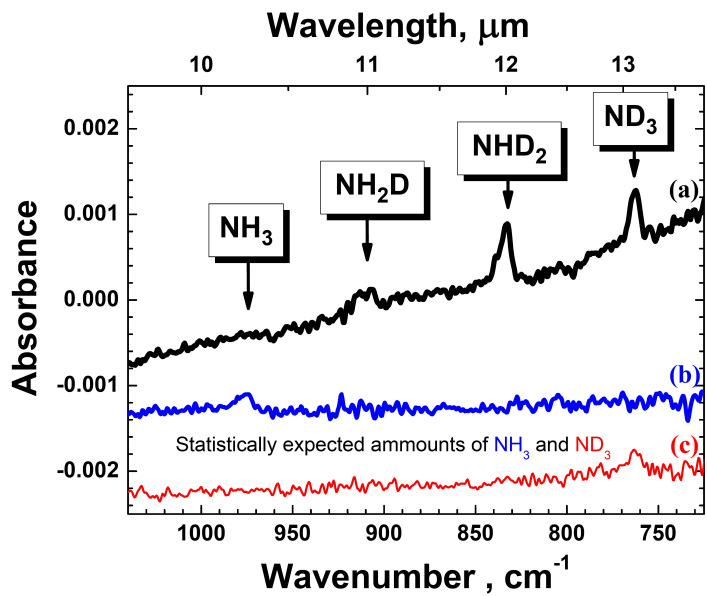

**Figure 2.** Three RAIR difference spectra: (a) co-deposition of N:(H+D):N$_2$:CO = 1:15:100:100 at 13 K (experiment 2.1); here, the H- and D-atom beams are prepared in the thermal cracking line by feeding the line with a mixture of H$_2$:D$_2$ = 1:1; (b) co-deposition of N:H:N$_2$:CO = 1:20:100:100 at 13 K (experiment 3.1); (c) co-deposition of N:D:N$_2$:CO = 1:20:100:100 at 13 K (experiment 3.2). The total N-atom fluence in (a) is 8 times higher than in (b) and (c). This factor 8 represents the weight of NH$_3$ or ND$_3$ molecules in a pure statistical distribution of final hydrogenation products assuming that H:D = 1:1 (*i.e.*, NH$_3$:NH$_2$D:ND$_2$H:ND$_3$ = 1:3:3:1).

Figure 3 shows the amount of ammonia isotopes formed as a function of the N-atom fluence. The uncertainties are large, but the derived data points hint for a linear growth for all the produced species. This is consistent with the assumption that the formation of all NH$_3$ isotopologues proceeds through subsequent H/D-atom addition to a single nitrogen atom, and no secondary processes like abstractions are involved (Hidaka et al. 2011, Fedoseev et al. 2014).

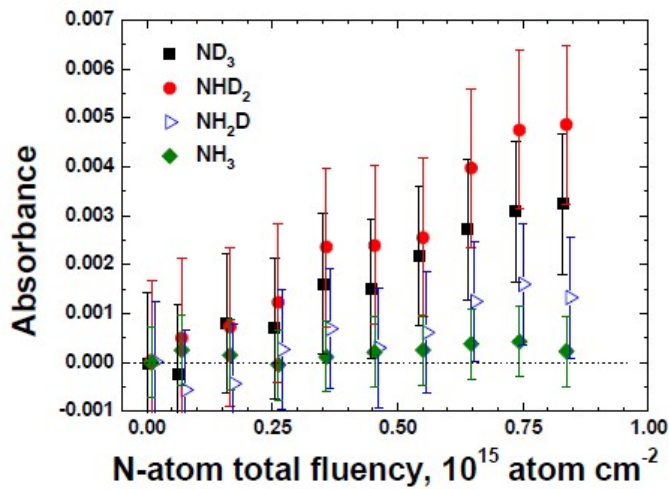

**Figure 3.** Co-deposition of N:(H+D):N$_2$:CO = 1:15:100:100 at 13 K (experiment 3.1). The absorbance of NH$_3$, NH$_2$D, ND$_2$H and ND$_3$ (taken from the integration of the $\nu_2$ mode) is shown as a function of the N-atom fluence.

### 3.3 Temperature dependency of deuterium enrichment of the produced $NH_{3-n}D_n$ isotopologues in N+H+D atom addition reactions

The N+H+D co-deposition experiments described in the previous section are here repeated for a number of different temperatures with the goal to study the thermal dependence of $NH_{3-n}D_n$ (with *n* = 0, 1, 2, 3) formation. The RAIR spectra obtained after co-deposition of $1.1 \cdot 10^{15}$ N atoms cm$^{-2}$ with a mixed H:D-atom beam at 13, 14, 15, 16, and 17 K are shown in Figure 4. Two conclusions can be derived from these plots. First, the formation of all four isotopologues is observed in the 13-15 K range, but drops below the detection limit between 16 and 17 K. This decrease of the total $NH_{3-n}D_n$ production confirms, once again, the conclusions in Fedoseev et al. (2014) that hydrogenation of N atoms in CO-rich ices takes place through the L-H mechanism. In the case of Eley-Rideal (E-R) or hot-atom mechanisms, no significant temperature dependence, specifically within such a small range, is expected. Secondly, the deviation of the observed signals from a statistical distribution in the amount of formed $NH_3$, $NH_2D$, $NHD_2$, and $ND_3$ in favour of D-substituted isotopologues remains constant, despite a gradual decrease of the total amount of products. In other words the observed preference in deuteration events over hydrogenation events is within our detection levels the same for all the tested temperatures.

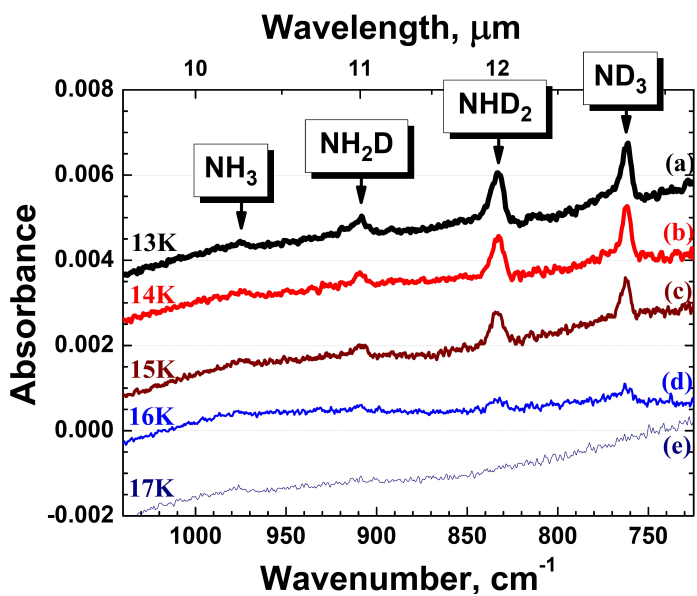

**Figure 4.** RAIR spectra obtained after co-deposition of N:(H+D):N$_2$:CO = 1:15:100:100 at five different temperatures: (a) 13 K, (b) 14 K, (c) 15 K, (d) 16 K, and (e) 17 K (see experiments 2.2-2.6). The mixed H- and D-atom beam is prepared in the thermal cracking line by feeding the line with a mixture of H$_2$:D$_2$ = 1:1.

### 3.4 Discussion

All four $NH_{3-n}D_n$ isotopologues are observed among the products of simultaneous co-deposition of N atoms with H and D atoms. This allows us to study for the first time the competition between hydrogenation and deuteration of N atoms in the solid phase at low temperatures. Previous work aimed at studying the hydrogenation of N atoms in a N$_2$ matrix (Hiraoka et al. 1995, Hidaka et al. 2011) as well as in CO-rich ices (Fedoseev et al. 2014). All three studies suggest that hydrogenation of nitrogen atoms takes place through subsequent H-atom addition to a single N-atom:

$$N + H \rightarrow NH \quad (1)$$
$$NH + H \rightarrow NH_2 \quad (2)$$
$$NH_2 + H \rightarrow NH_3 \quad (3).$$

When D atoms are introduced simultaneously with H atoms, it is logical to assume that these also will participate in competing deuteration reactions, following the same chemical pathway:

$$N + H/D \rightarrow NH_{1-n}D_n \quad (n = 0, 1) \quad (4)$$
$$NH_{1-n}D_n + H/D \rightarrow NH_{2-n}D_n \quad (n = 0, 1, 2) \quad (5)$$
$$NH_{2-n}D_n + H/D \rightarrow NH_{3-n}D_n \quad (n = 0, 1, 2, 3) \quad (6).$$

As stated before, in the case that there is no chemical preference for H- or D-addition reactions, then for a H:D = 1:1 mixture, the final isotopologue distribution follows a statistical weighting of $NH_3:NH_2D:ND_2H:ND_3 = 1:3:3:1$. However, the actually observed distribution deviates, and is determined as 0.4:2.1:3.5:2, in favour of a higher (1.7 times) deuteration efficiency compared to hydrogenation. To determine the exact process responsible for this enrichment, we have considered different possibilities: experimental artefacts, specifically, (*i*) deviations from the H:D = 1:1 ratio in the mixed atom beam fluxes in favour of D atoms; or a physically and chemically different behaviour, *i.e.*, (*ii*) differences in accretion rates (sticking probabilities) of H and D atoms, (*iii*) differences in desorption and diffusion rates of the adsorbed H and D atoms, and (*iv*) other competing reactions involved, like for example atom abstraction reactions. These possibilities are discussed separately.

(*i*) *Lower H-atom flux over D-atom flux in the mixed beam*.
As we described before, a $H_2:D_2 = 1:1$ gas mixture is used as a feeding gas for the thermal cracking line to obtain a mixed H/D-atom beam. The difference in mass and bond energy of both precursor species may yield an $H:D:H_2:D_2:HD$ beam reflecting a deviating distribution, causing experiments to start from (unknown) initial conditions. Therefore, to evaluate the exact H:D ratio in the atom beam, co-deposition experiments are performed, where mixed H/D atoms are co-deposited with large overabundances of $O_2$ molecules and - as discussed in the experimental section - the amount of $HO_2$ and $DO_2$ formed through barrierless reactions can be directly linked to the H/D ratio. The test shows that the same amount of $HO_2$ and $DO_2$ is formed in all experiments at all tested temperatures. The temperature independence of the obtained results is in strong support of an E-R mechanism for $HO_2$ and $DO_2$ formation. Therefore, assuming the same probability for H and D atoms to react with $O_2$ molecules upon encounter, the H:D-atom ratio at the ice surface is equal to 1 and cannot explain the observed deuterium enrichment in N+H+D co-deposition experiments.

(*ii*) *Different sticking probabilities for H and D atoms at the surface of the ice*.
Several studies have been devoted to the investigation of the sticking coefficients of H atoms to cryogenically cooled surfaces as a function of the translational energy of the H atoms (for an overview see Watanabe et al. 2008). Buch & Zhang (1991) performed molecular dynamic simulations showing that D atoms have a higher sticking probability (to $H_2O$ ice) than H atoms for all studied temperatures (50-600 K). For instance, a 300 K D-atom beam has a calculated sticking probability to water ice 2.5 higher than H atoms. Unfortunately, to date, no experimental data is available on a mixed $CO:N_2$ ice surface and a 300 K estimated D/H-atom beam temperature. However, following the simulations, we expect a higher surface density of D atoms (w.r.t. H atoms) on the surface of the ice, consistent with the observed results. It should be noted, though, that this may not be the full story. The nose-shaped quartz pipe is used to collisionally quench 'hot' atoms. Therefore, it is not necessarily true that H and D atoms have identical translational energies at the ice surface; both species are light and this makes efficient collisional quenching even more challenging. Finally, the factor 2 difference in mass may affect the resulting distribution of translational energies, further changing the sticking coefficients for H and D atoms. Bottom line, the sticking probability for H and D atoms at cold surfaces seems to be a key parameter to explain the experimentally observed deuterium fractionation.

(*iii*) *Difference in desorption and diffusion rates for H and D atoms.*
If D atoms have a higher binding energy to CO-ice than H atoms, this can affect the system in two opposite ways. On one hand, a lower binding energy of H atoms means that they have a higher probability to desorb from the surface of the ice before a reaction with other species takes place. On the other hand, a lower binding energy means that, for the studied temperature range, thermal hopping of H atoms is to occur more frequently than thermal hopping of D atoms. If quantum tunnelling is responsible for the diffusion of H/D atoms, this further enhances the mobility of H atoms and the probability to find an N-atom to yield $NH_{1-n}D_n$ ($n$ = 0, 1) and $NH_{2-n}D_n$. ($n$ = 0, 1, 2). To verify this hypothesis, a set of experiments as described in section 3.3 was performed. No significant difference upon sample temperature variation is found up to 17 K, where none of the reaction products could be detected. In the case that either a difference in desorption or diffusion rate would be responsible for the observed deuterium enrichment, one would expect to find a significant change in the $NH_3:NH_2D:ND_2H:ND_3$ distribution with temperature from maximum production to non-production of ammonia isotopologues. This is not observed and hints for the conclusion that different diffusion rates are not case determining. This is also consistent with the small isotopic difference between the diffusion of H and D atoms on amorphous water ice as observed by Hama et al. (2012).

(*iv*) *Competing reaction channels*.
Taking into account an overabundance of H/D atoms compared to N atoms on the surface of the ice, the main competing reaction channels are expected to be:

$$H + H \rightarrow H_2 \qquad (7)$$
$$D + D \rightarrow D_2 \qquad (8)$$
$$H + D \rightarrow HD \qquad (9).$$

If two separate systems would be studied (*i.e.*, N+H and N+D), the rates of the competing barrierless reactions (7) and (8) would affect the rates of $NH_3$ and $ND_3$ formation through the consumption of the available H and D atoms, and would decrease the corresponding H and D surface densities. This is especially important if the $NH_3$ and $ND_3$ formation mechanism would involve activation barriers, slowing down reactions in comparison to reactions (7) and (8). However, this is not the case. In addition, in our system, where H and D atoms are co-deposited, reaction (9), which decreases surface densities of H and D atoms in an equal way, should dominate. Our data cannot conclude precisely on the difference in surface H- and D-atom densities during the experiments, and knowledge of the activation barrier for H and D diffusion and desorption is required to build a precise model to investigate it. Other possible competing reactions that should be mentioned here are abstraction reactions. Although reaction

$$NH_3 + D \rightarrow NH_2 + HD \qquad (10)$$

can be excluded experimentally from the list (see section 3.1), reactions

$$NH_2 + D \rightarrow NH + HD \qquad (11)$$
$$NH + D \rightarrow N + HD \qquad (12)$$

and similar reactions, including deuterated isotopologues + H atoms instead of D atoms, should not be disregarded. Also the abstraction reactions:

$$NH_2 + H_2/D_2 \rightarrow NH_3/NH_2D + H/D \qquad (13)$$
$$NH + H_2/D_2 \rightarrow NH_2/NHD + H/D \qquad (14)$$
$$N + H_2/D_2 \rightarrow NH/ND + H/D \qquad (15)$$

should be considered. Both reactions (11) and (12) are expected to have activation barriers. Similar reactions involving only H atoms show barriers of 6200 K for reaction (11) and 400 K for reaction (12) (Ischtwan & Collins 1994, Poveda & Varandas 2005). Reactions (13) - (15) are expected to have activation barriers as well and are endothermic (see Hidaka et al 2011 and references therein). Since reactions (4)-(6) are all barrierless, they should proceed much faster than reactions (11)-(15). Therefore, we assume that reactions (11)-(15) do not contribute significantly to the formation of ammonia and its isotopologues under our experimental conditions.

After discussing these arguments, we consider the difference in sticking probabilities of H and D atoms to the surface of CO-rich ices (resulting in a higher surface density of D atoms over H atoms) as the main reason for the observed deuterium enrichment of ammonia isotopologues produced by surface hydrogenation/deuteration of N atoms at low temperatures.

## 4 ASTROCHEMICAL IMPLICATIONS AND CONCLUSIONS

The observed high deuterium fractionation in prestellar cores (high densities $n \geq 10^6$ cm$^{-3}$ and low temperatures $T \leq 10$ K) is the result of a combination of gas-phase and surface reactions. Under molecular cloud conditions, the D/H ratio of molecules is found orders of magnitude higher (between 0.02-0.09 for DNC/HNC) than the elemental abundance of D/H = $1.5 \times 10^{-5}$ (Linsky 2003, Hirota et al. 2003). It is well established that this enhancement is largely due to exothermic exchange gas-phase reactions involving $H_3^+$ (Oka 2013, Millar et al. 1989, Roberts et al. 2004). A process that can counterbalance the deuterium enrichment of gas-phase species is the reaction between $H_2D^+$ and CO. However, for cloud densities higher than a few $10^5$ cm$^{-3}$, timescales for collisions of CO with grains become so short that most of the gaseous CO is depleted from the gas to form a layer of pure CO ice on the grains. This so-called 'catastrophic' freeze-out of CO, observed directly through infrared ice-mapping observations (Pontoppidan 2006) and indirectly through the lack of gas-phase CO and other molecules in dense regions (Bergin et al. 2002, Caselli et al. 1999), causes a rise in gaseous $H_2D^+$ and deuterated molecules. Moreover, electron recombination of $H_2D^+$ enhances the abundance of D atoms, which then can participate in surface reactions on dust grains to form deuterated ice (Aikawa 2013).

Recent laboratory experiments proved that surface reactions involving deuterium - including hydrogen abstraction reactions - lead to the deuterium enrichment of interstellar ices. For instance, the deuteration of solid $O/O_2/O_3$ induces the formation of deuterated water ice (*e.g.*, Dulieu et al. 2010, Ioppolo et al. 2008, Romanzin et al. 2011). However, the formation of $H_2O/HDO$ through $OH+H_2/D_2$ and $H_2O_2+H/D$ (Oba et al 2012, Oba et al 2014) shows a preference for hydrogenation that has been explained by a higher quantum tunneling efficiency. On the other hand, Nagaoka at al. (2005), Nagaoka (2007), and Hidaka at al. (2009) demonstrated that hydrogen atoms can be abstracted from methanol and its isotopologues and substituted by D atoms upon D-atom exposure of solid $CH_3OH$, $CH_2DOH$, and $CHD_2OH$.

Triply deuterated ammonia ($ND_3$) has been observed in dark clouds with an $ND_3/NH_3$ abundance ratio of $\sim 8 \cdot 10^{-4}$, which implies an enhancement of more than 10 orders of magnitude over the purely statistical value expected from the abundance of deuterium in the interstellar medium (Lis et al. 2002). The deuterium enrichment of ammonia can occur both in the gas phase and in the solid phase (Rodgers & Charnley 2001). The gas-phase synthesis of ammonia in cold dense clouds occurs through a sequence of ion-molecule reactions that start with the fragmentation of $N_2$ and formation of $N^+$ ions through the reaction with $He^+$, which is formed by cosmic-ray ionization of He. Successive reactions with $H_2$ end with the formation of the $NH_4^+$ ion, and dissociative recombination with electrons finally yields $NH_3$ as the dominant product (Vikor et al. 1999, Öjekull et al. 2004, Agúndez & Wakelam 2013). When deuterated species are involved in the process, triply deuterated ammonia is then formed through the dissociative recombination of $NHD_3^+$. In their gas-phase chemical model, Lis et al. (2002) were able to reproduce the observed abundances of deuterated ammonia only when a non-statistical ratio for the dissociative recombination reaction forming $ND_3$ was used. An alternative route is the deuterium fractionation on the grains, as studied here.

In the solid phase, the formation of ammonia is a radical-radical process (reactions (1)-(3)) that occurs at low temperatures (<15 K). Previous laboratory work confirms that these surface reactions are nearly barrierless (Hiraoka et al. 1995, Hidaka et al. 2011, Fedoseev et al. 2014). Therefore, one would expect the deuterium fractionation of ammonia to reflect the atomic D/H ratio in the accreted gas. Disregarding any other surface reaction and assuming that the probability for accretion of N atoms and reaction with D atoms is $p_D$, then the probability for reaction with H atoms is (1- $p_D$). The expected fractionation for $NH_2D/NH_3$ is $3p_D/(1- p_D)$, where the factor 3 accounts for the three chances to deuterate ammonia, and is $3p^2_D/(1- p_D)^2$ for $ND_2H/NH_3$ (Tielens 2005). Although this simple calculation strongly depends on the local cloud conditions, it is based on a statistical distribution of the $NH_{3-n}D_n$ isotopologues and does not fully reflect the observed gas-phase abundances for the $NH_{3-n}D_n$ isotopologues in dense cold clouds (Rouef et al. 2000, Loinard et al. 2001).

Our laboratory results indicate that the simultaneous addition of H and D atoms to N atoms on a cold surface leads to the formation of all the deuterated isotopologues of ammonia with a distribution that is non-statistical at low temperatures and that leads to a higher deuterium fractionation of ammonia ice. Our experiments are performed in CO-rich ices to resemble the conditions found in dense cold cores, where CO freezes-out onto dust grains and the D/H ratio increases. The use of a CO matrix in our experiments also helps to overcome the IR spectral broadening that occurs in polar ices and that induces spectral confusion. Moreover, the D/H ratio (~1:1) chosen for our experiments is somehow representative to the densest prestellar cores and, at the same time, simplifies the data analysis. Under our experimental conditions, every deuteration event has a probability of at least a factor 1.7 higher to occur over a regular hydrogenation event, independently from the surface temperature. A higher sticking probability of D atoms over H atoms to the surface of the ice can explain our experimental findings. In this scenario, the surface D/H ratio is higher than the already enhanced gas-phase D/H ratio as in dense cores, because D atoms have a higher binding energy to the surface than H atoms. This will further increase the deuterium fractionation of species that are formed in the solid phase and then later released into the gas-phase. Therefore, our results show that the deuterium fractionation of species in the solid phase is potentially a more important process than previously considered.

This conclusion can be extrapolated to other chemical systems than simply ammonia ice, where isotopologues are formed by a series of competing and barrierless H/D-atom addition reactions to a single atom. For instance, we expect that the competition between hydrogenation and deuteration of $CH_4$ leads to an enhancement of deuterated species. Our H- and D-atom beam has a temperature of 300 K, as opposed to the temperature of H and D atoms in dark cloud regions that is roughly an order of magnitude lower. Buch & Zhang (1991) reported that the sticking probability of 300 K deuterium atoms to the surface of an amorphous $H_2O$ cluster is 2.5 higher than the sticking probability of hydrogen atoms. The same ratio goes down to 1.42, when the D and H atoms are at 50 K. Moreover, different surface properties, *i.e.*, polar ($H_2O$-rich) *versus* non-polar (CO-rich) ice, can potentially affect the sticking probabilities of H and D atoms as well. Therefore, we expect that complementary systematic experiments performed with H- and D-atom beams at different kinetic temperatures and in both polar and non-polar ice analogues will be pivotal to determine the role of ice grain chemistry in the interstellar deuterium fractionation of molecules like ammonia. The present study already shows that surface reactions clearly can contribute to the observed gas phase abundances.

**ACKNOWLEDGMENTS**

We are grateful to Herma Cuppen (Nijmegen) for fruitful discussions during the preparation of the manuscript. This research was funded by the European Community's Seventh Framework Programme (FP7/2007-2013) under grant agreement no. 238258 (LASSIE), the Netherlands Research School for Astronomy (NOVA), Netherlands Organization for Scientific Research (NWO) through a VICI grant and the European Research Council (ERC-2010-StG, Grant Agreement No. 259510-KISLMOL). Support for S.I. from the Niels Stensen Fellowship and the Marie Curie Fellowship (FP7-PEOPLE-2011-IOF-300957) is gratefully acknowledged.